**Competition and coexistence of antiferromagnetism and superconductivity in underdoped $Ba(Fe_{0.953}Co_{0.047})_2As_2$**

D. K. Pratt, W. Tian, A. Kreyssig, J. L. Zarestky, S. Nandi, N. Ni, S. L. Bud'ko, P. C. Canfield, A. I. Goldman, and R. J. McQueeney
*Department of Physics and Astronomy and Ames Laboratory, Iowa State University, Ames, IA 50011 USA*

## Abstract

Neutron and x-ray diffraction studies show that the simultaneous first-order transition to an orthorhombic and antiferromagnetic (AFM) ordered state in $BaFe_2As_2$ splits into two transitions with Co doping. For $Ba(Fe_{0.953}Co_{0.047})_2As_2$, a tetragonal-orthorhombic transition occurs at $T_S$ = 60 K, followed by a second-order transition to AFM order at $T_N$ = 47 K. Superconductivity (SC) occurs in the orthorhombic state below $T_C$ = 15 K and coexists with AFM. Below $T_C$, the static Fe moment is reduced and a 4 meV spin gap develops indicating competition between coexisting SC and AFM order.

In the $R$FeAsO and $A$Fe$_2$As$_2$ compounds ($R$=rare-earth, $A$=alkali earth), superconductivity (SC) appears as the transition to an orthorhombic and antiferromagnetic (AFM) ordered state is suppressed by chemical doping.[1-5] Detailed studies of the evolution of structural, magnetic, and SC properties as a function of chemical doping have revealed more behavior for compositions close to the boundary between (or coexistence region of) AFM and SC phases. In some cases, such as CeFeAsO$_{1-x}$F$_x$,[6] the phases appear to be mutually exclusive. However, in other cases such as Ba$_{1-x}$K$_x$Fe$_2$As$_2$ [7], Ba(Fe$_{1-x}$Co$_x$)$_2$As$_2$,[4, 5] and SmFeAsO$_{1-x}$F$_x$ [8], regions of the phase diagram indicate a coexistence of AFM and SC. Coexistence of AFM and SC has been observed in other magnetic superconductors such as the Chevrel phases ($R$Mo$_6$S$_8$),[9] borocarbides ($R$Ni$_2$B$_2$C),[10] and ruthenates (RuSr$_2$GdCu$_2$O$_8$) [11] where rare-earth magnetic ordering occurs independently of SC. However, in some compounds, both magnetism and SC evolve from the same electronic (conduction) bands. In these exceptional systems, such UPt$_3$ [12, 13] and UNi$_2$Al$_3$,[14] neutron and resonant x-ray diffraction measurements have shown clear coupling between AFM and SC order parameters. The FeAs superconductors belong to the latter category, since both AFM and SC originate from the Fe d bands. In this Letter, we present neutron diffraction measurements on Ba(Fe$_{0.953}$Co$_{0.047}$)$_2$As$_2$ that show a strong decrease in the AFM order parameter below $T_C$. In addition, gapless spin wave excitations observed above $T_C$ become gapped in the SC state. Taken together, the results provide strong evidence supporting competition between homogeneous and coexisting AFM and SC phases.

The series of Co-doped $Ba(Fe_{1-x}Co_x)_2As_2$ has been studied systematically by heat capacity, magnetization, resistivity,[4, 5] and thermal expansion [15] measurements on single-crystals grown in self-flux. With the addition of Co, the temperature of the first-order transition from the tetragonal (*T*) phase to the orthorhombic (*O*) and AFM ordered phase decreases and the transition either broadens or splits into separate transitions. Although it is suspected that the transition has structural and/or magnetic character, no microscopic determination of these phases has been reported until now. Nonetheless, evidence of the purported magnetic/structural transitions can be observed up to $x = 0.06$. [4, 5] At compositions above $x = 0.03$, SC first appears and exhibits a region between $x \sim 0.03 – 0.06$ where it is conjectured that the SC phase apparently occurs in the presence of the orthorhombic AFM phase (called underdoped compositions). [4, 5] Beyond $x = 0.06$, the magnetic/structural transition seems completely suppressed and SC transition temperature reaches a maximum of 23 K for $x = 0.074$. [4, 5]

Figures 1 (a) and 1(b) show the magnetization and resistivity data for an underdoped sample with $x = 0.047$. The data clearly shows a superconducting transition at $T_C = 15$ K. Using the criteria in Ref. [4], two higher temperature anomalies corresponding to the split transition are also identified. Upon cooling from high temperatures, the magnetization and resistivity undergo a change in slope near 60 K with temperature derivatives in figs. 1(a) and 1(b) showing the anomaly more clearly. This is followed by another transition near 47 K. In analogy with the *R*FeAsO series of compounds where the magnetic and structural transitions are split,[16] it is speculated that the strong resistivity anomaly at 60 K is associated with the *T-O* transition and the lower one (47 K) to AFM ordering.

We undertook neutron and x-ray diffraction experiments to determine if the transitions separate and, if so, to identify the structural and magnetic phases in the underdoped region. It is also interesting to study how the AFM ordering, if present, evolves into the SC phase. Diffraction experiments were performed on single-crystals of underdoped $Ba(Fe_{0.953}Co_{0.047})_2As_2$ that were grown under identical conditions as those samples used in the bulk measurements in figs. 1 (a) and (b) and described in detail in Ref. [4]. Neutron diffraction measurements were performed on the HB1A diffractometer at the High Flux Isotope Reactor at Oak Ridge National Laboratory on a sample weighing approximately 700 mg and having a crystal mosaic width of <0.3 degrees. The experimental configuration was 48'-40'-40'-136' with $E_i$ = 14.7 meV. The sample was aligned in the tetragonal [*HHL*] plane and mounted in a closed-cycle refrigerator for low temperature studies. The temperature dependence was studied at several nuclear Bragg peak positions and at $\mathbf{Q}_{\mathbf{AFM}} = (1/2,1/2,L = odd)$ positions corresponding to the AFM ordering in the parent $BaFe_2As_2$ compound.[17]

Figure 1(c) shows the evolution of the integrated intensity of the (220) nuclear reflection with temperature. The (220) intensity starts to increase at about 80 K and grows gradually over a range of 20 K before increasing sharply at $T_S$ = 60 K. This increase in intensity is ascribed to extinction release that occurs due to the formation of *O* twin domains at the structural transition. While the resolution of the neutron diffraction experiments were insufficient to determine the *O* splitting, we were able to confirm the orthorhombicity by performing high-energy single-crystal x-ray diffraction experiments using MUCAT sector 6-ID-D at the Advanced Photon Source at Argonne National Laboratory with an incident photon energy of 99.5 keV. X-ray diffraction measurements of the (220) reflection above and below $T_S$ (Fig. 2) clearly show a very small *O* splitting ($(a-b)/(a+b)$ = 0.12%) and twinning in the $x = 0.047$ sample. Given the full penetration of the x-rays, the results suggest that single phase *O* structure exists throughout the crystal.

To confirm the single phase *O* structure, a slightly lower composition of $x = 0.038$ was studied. This composition also shows a split transition and superconductivity, and the larger *O* splitting of 0.2% allows for clearer separation of twin reflections. Fig 2 shows the temperature evolution of the (1,1,10) reflection for $x$ =0.038 measured using a Rigaku rotating Cu-anode RU-300 system. The *O* splitting grows continuously below 76 K and there does not appear to be any remaining *T*-phase. While the evolution of the (220) intensity in the $x = 0.047$ neutron data is very sensitive to the structural transition, it arises from the formation of extrinsic twin domains and cannot be considered as the order parameter of the *T-O* structural transition. Despite the lack of strong evidence for coexisting *T* and *O* phases (fig 2) and thermal hysteresis (fig 1(c)) we cannot determine if the *T-O* transition for the $x = 0.047$ composition is first- or second-order.

Fig. 1(c) shows intensity gradually appearing at the (1/2,1/2,1) magnetic Bragg position below $T_N = 47$ K. The magnetic wavevector is identical to that for $BaFe_2As_2$ [17] indicating that the magnetic structure is likely the same AFM "stripe" structure observed in all of the AFM ordered Fe-As materials. The AFM squared order parameter was obtained from the integrated intensity of the (1/2,1/2,1) peak. Below $T_N$, the integrated intensity can be fit to the form $(T_N - T)^{-2\beta}$ with exponent $2\beta = 0.6$ and consistent with a second-order transition. An extrapolated zero-temperature magnetic moment of 0.2 $\pm$ 0.1 $\mu_B$ was estimated by comparison of the (1/2,1/2,1) intensity to nuclear Bragg intensities and also to magnetic intensities of $CaFe_2As_2$ under similar experimental conditions. The moment is significantly reduced compared to the $x = 0$ moment of 0.87 $\mu_B$, similar to the reduction observed in other doped FeAs compounds.[6, 18] The evolution of the nuclear and magnetic intensities shown in Fig 1(c) confirms that the single first-order transition in $BaFe_2As_2$ has split into two separate transitions with Co addition with the lower, magnetic, transition appearing to be second-order in nature. Similar to the *R*FeAsO compounds,[16] the structural transition occurs first upon cooling, followed by AFM ordering. Figure 3 identifies these states in the phase diagram for $Ba(Fe_{1-x}Co_x)_2As_2$.

The integrated AFM intensity is substantially reduced below $T_C$, as shown in Fig. 1(c). Meanwhile, the intensity ratio of the (1/2,1/2,1) and (1/2,1/2,3) reflections is unchanged (fig. 4(a)) indicating that the moment direction and magnetic symmetry are unaffected by the onset of $T_C$. The partial suppression of (1/2,1/2,1) intensity shown in fig 4(a) therefore suggests a reduction in the average static Fe moment below $T_C$. In addition, inelastic energy scans at (1/2,1/2,1) (fig. 4(b)) show that the low energy magnetic spectral weight is also suppressed. Above $T_C$, the excitations are gapless and constant energy cuts at 2.5 meV along the [110] direction reveal a sharp peak (fig 4(b), inset) consistent with steep spin wave excitations.[19] Below $T_C$, the intensity below 4 meV (and spin wave peak at 2.5 meV) are suppressed demonstrating that a gap forms in the spin wave excitations. The gapping of AFM spin excitations below $T_C$ indicates a direct coupling between AFM and SC and is strong evidence in support of the homogeneous coexistence of AFM and SC (see below). The behavior of the low energy spin excitations is similar to the development of a gap and resonance-like feature below $T_C$ in the optimally doped superconductor with $x = 0.1$.[20] However, at the optimal composition there is no long-range AFM order and SC gaps the spectrum of short-ranged spin fluctuations that are much broader in momentum space.

We now address the question of homogeneity and coexistence of the AFM, SC, and *O* phases in underdoped compositions. Evidence from transport and thermodynamic measurements [4] and magnetic optical imaging of Meissner flux expulsion [21] supports homogeneous SC. X-ray measurements in fig. 2 find only the twinned *O*-structure in the SC phase, with no observable *T*-phase. Thus, crystals appear to have crystallographic homogeneity from which one can conclude that SC occurs in the *O*-structure. This is supported by the observation of a distinct difference in the anisotropy of $H_{C2}$ in underdoped (orthorhombic) and optimally doped (tetragonal) samples.[4] Neutron diffraction alone cannot provide direct evidence for homogeneity of AFM. However, as discussed above, gapping of the low energy spin excitations together with the reduction of the static moment provides strong evidence that AFM and SC coexist homogeneously and are in competition with one another.

Competition between AFM and SC is natural when one considers that both originate from the multiple Fe conduction bands that cross the Fermi level. Fermi surface nesting giving rise to AFM or spin density wave (SDW) ordering will gap only part of the Fermi surface. On the other hand, SC will (for example) gap the entire Fermi surface for an *s*- wave gap, or form line nodes for a *d*-wave gap. The (repulsive) interaction between SDW and SC order parameters then arises from competition over the shared electronic density-of-states common to both gaps, regardless of the microscopic origins of SC. This has been demonstrated using two-band itinerant model for the iron arsenides where competing SDW and *s*+ SC are shown to coexist over a range of doping.[22] For a more general case of *s*-wave [23] or *d*-wave SC [24], the competition causes a reduction sublattice magnetization below $T_C$. In $Ba(Fe_{0.953}Co_{0.047})_2As_2$, the observed reduction in the ordered moment below $T_C$ is substantially larger than that observed $UPt_3$ [12] or $UNi_2Al_3$[14], demonstrating an unusually strong interaction between AFM and SC in the iron arsenides.

## ACKNOWLEDGMENTS

The authors would like to thank J. Schmalian, D. N. Argyriou, and R. Prozorov for valuable comments and D. S. Robinson for excellent support of the x-ray study. Work at the Ames Laboratory and the MUCAT sector is supported by the U.S. Department of Energy Office of Science under Contract No. DE-AC02 07CH11358. Use of the Advanced Photon Source was supported by US DOE under Contract No. DE-AC02-06CH11357.

FIGURE CAPTIONS

FIG. 1. (a) The magnetization (dots) and its temperature derivative (line), and (b) the resistivity and its temperature derivative for single-crystal $Ba(Fe_{0.953}Co_{0.047})_2As_2$ as a function of temperature. (c) The integrated intensity of the (220) nuclear reflection (circles) and the (1/2,1/2,1) magnetic reflection (squares) as a function of temperature. Hollow symbols indicate warming and filled symbols cooling. The solid line shows the power law fit to the magnetic order parameter. Vertical lines through all three panels indicate the structural ($T_S$), magnetic ($T_N$), and superconducting ($T_C$) transitions.

FIG. 2. (Right) Images of the x-ray diffracted intensity of the (220) peak for $x = 0.047$ showing a single spot above $T_S$ and two spots below $T_S$ due to orthorhombic splitting. (Left) The temperature evolution of the (1,1,10) reflection for $x = 0.038$, showing *T-O* phase transition.

FIG. 3. Phase diagram for $Ba(Fe_{1-x}Co_x)_2As_2$ showing paramagnetic tetragonal (T), paramagnetic orthorhombic (O), AFM ordered orthorhombic (AFM O), and superconducting (SC). AFM O and SC phases coexist between $x = 4$-6%. The vertical line shows the position of the $x = 0.047$ sample studied here. The data points for the phase lines have been taken from Ref. [4].

FIG. 4. (a) The reduction in intensity of (1/2,1/2,1) peak compared to the fit of the magnetic order parameter to a power law (shown in fig. 1(c)) and the intensity ratio of (1/2,1/2,1) and (1/2,1/2,3) as a function of temperature. (b) Energy scan at (1/2,1/2,1) for T = 25 K (empty) and 5 K (solid). (c,inset) Scans along [110] at 2.5 meV for 25 K and 5 K.  Lines are guided to the eye.

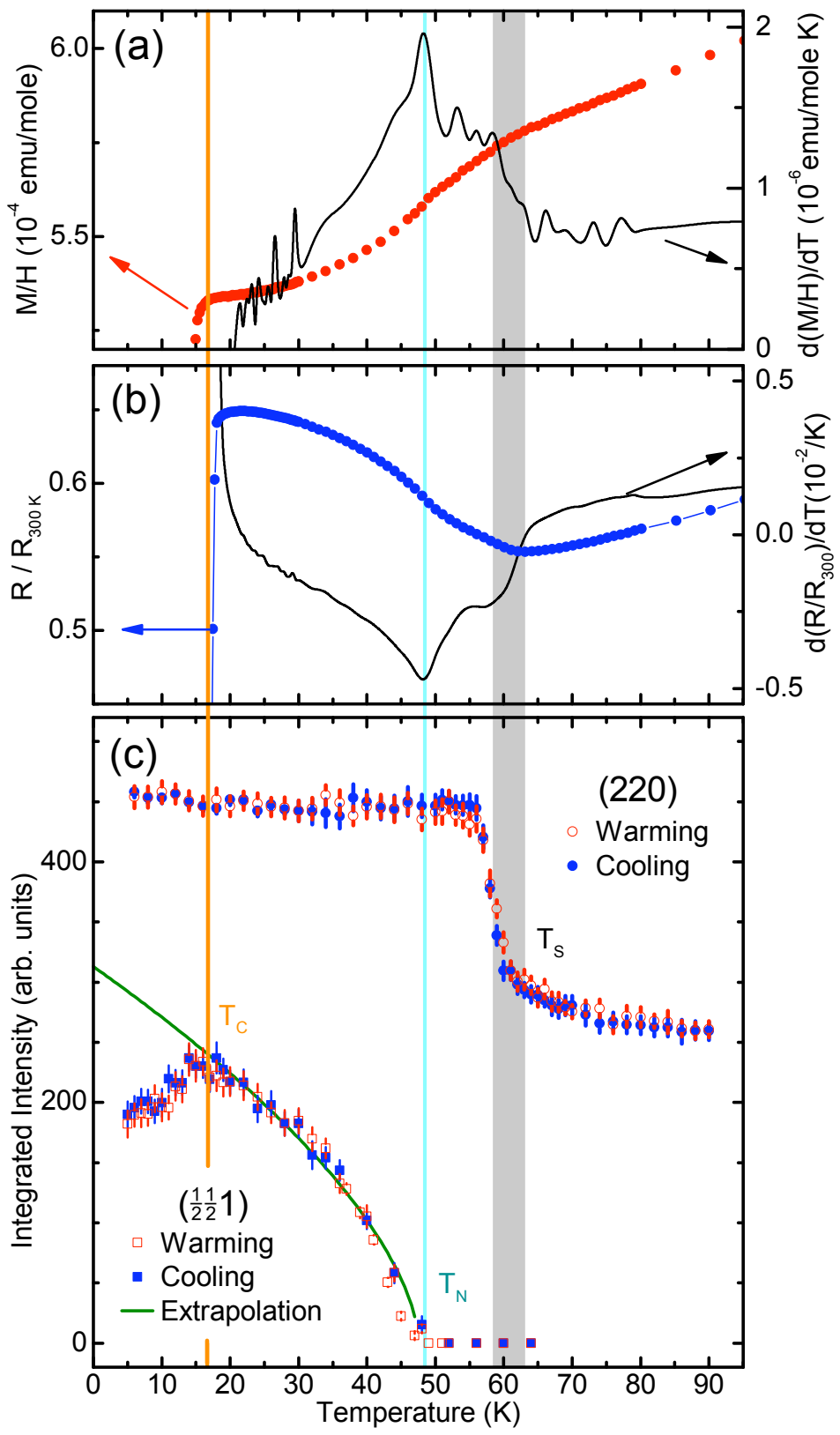

FIG. 1.

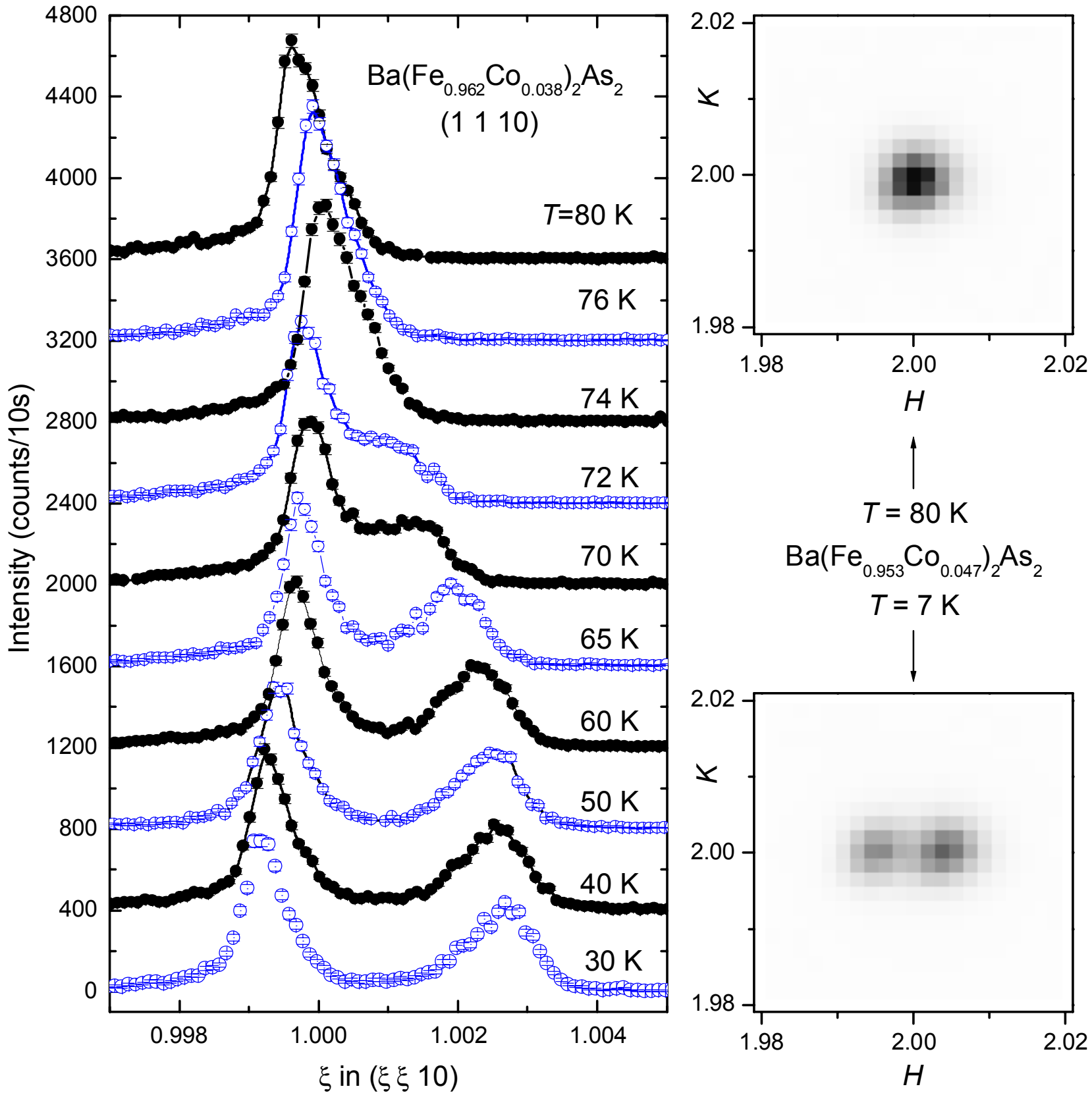

FIG. 2.

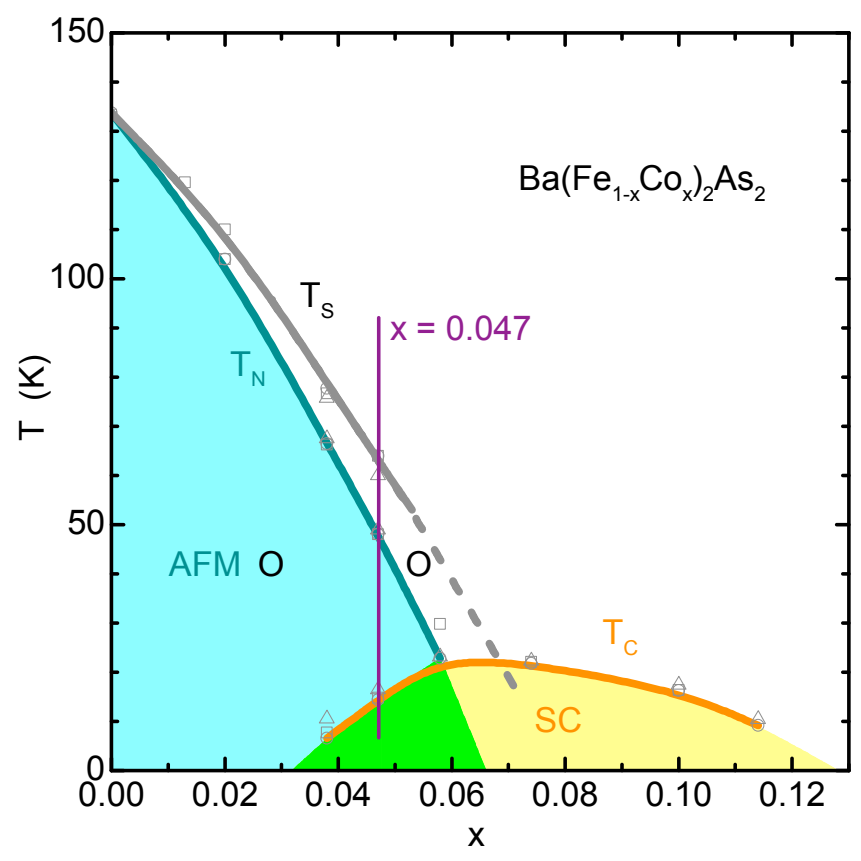

FIG. 3.

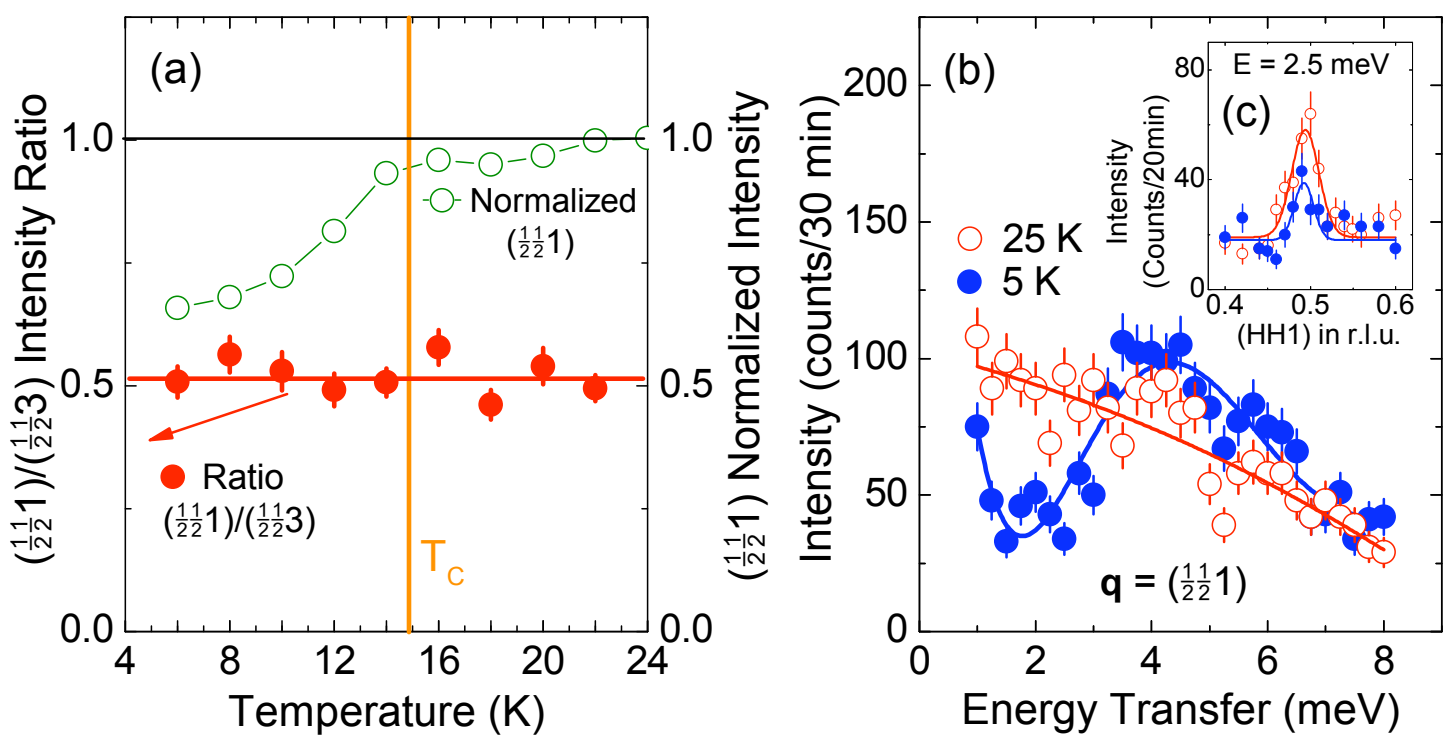

FIG. 4